\documentclass{pnastwo}

\usepackage{amssymb,amsfonts,amsmath,color}
\usepackage[dvips]{graphicx}
\usepackage{axodraw2,color}

\def\de{\mathrm d}

\let\a=\alpha \let\b=\beta  \let\g=\gamma  \let\d=\delta \let\e=\varepsilon
     \let\l=\lambda
\let\m=\mu                
 \let\t=\tau    
   
\let\G=\Gamma \let\D=\Delta  \let\L=\Lambda 
       
 \let\ee=\epsilon \let\r=\rho \let\th=\theta
\let\io=\infty  
\def\ie{{i.e. }}

\def\ba{\begin{align}}
\def\ea{\end{align}}

 \def\BB{{\cal B}}
  
\def\AA{{\cal A}}

\def\to{\rightarrow}
\def\la{\left\langle}
\def\ra{\right\rangle}

\def\de{\mathrm d}

\newcommand{\beq}{\begin{equation}} \newcommand{\eeq}{\end{equation}}
 
\contributor{Submitted to Proceedings
of the National Academy of Sciences of the United States of America}
\url{www.pnas.org/cgi/doi/10.1073/pnas.XXXXXXXXXX}
\copyrightyear{2012}
\issuedate{Issue Date}
\volume{Volume}
\issuenumber{Issue Number}

\begin{document}

\title{Quantitative field theory of the glass transition}

\author{Silvio Franz\affil{1}{Laboratoire de Physique Th\'eorique et Mod\`eles
    Statistiques, CNRS et Universit\'e Paris-Sud 11,
UMR8626, B\^at. 100, 91405 Orsay Cedex, France},
Hugo Jacquin\affil{2}{Laboratoire Mati\`ere et Syst\`emes Complexes, UMR 7057,
CNRS and Universit\'e Paris Diderot -- Paris 7, 10 rue Alice Domon et L\'eonie 
Duquet, 75205 Paris cedex 13, France},
Giorgio Parisi\affil{3}{Dipartimento di Fisica,
Sapienza Universit\'a di Roma,
INFN, Sezione di Roma I, IPFC -- CNR,
P.le A. Moro 2, I-00185 Roma, Italy},
Pierfrancesco Urbani\affil{3}{}\affil{1}{},
\and Francesco Zamponi\affil{4}{LPT,
Ecole Normale Sup\'erieure, UMR 8549 CNRS, 24 Rue Lhomond, 75005 France}
}

\contributor{Submitted to Proceedings of the National Academy of Sciences
of the United States of America}

\maketitle

\begin{article}

\begin{abstract} 
We develop a full microscopic replica field theory of the dynamical transition in glasses. 
By studying the soft modes that appear at the dynamical temperature we obtain an effective 
theory for the critical fluctuations. 
This analysis leads to several results: we give expressions for the mean field critical exponents, 
and we study analytically the critical behavior of a set of four-points correlation functions 
from which we can extract the dynamical correlation length. 
Finally, we can obtain a Ginzburg criterion that states the range of validity of our analysis.
We compute all these quantities within the Hypernetted Chain Approximation (HNC) for the Gibbs free energy
and we find results that are consistent with numerical simulations.
\end{abstract}


\subsection{Introduction}

Dynamical heterogeneities in structural glasses have been the object of
intensive investigations in the last 15 years~\cite{BBBCS11}. 
The early Adams--Gibbs theory of glass formation
was based on the concept of cooperatively rearranging
regions, whose size becomes larger and larger when the glass
region is approached. 
Such large cooperatively rearranging regions imply the
existence of dynamical heterogeneities characterized by a large
correlation length.
Large scale dynamical heterogeneities are expected to be present in any
framework where glassiness is due to collective effects: they are
indeed the smoking guns for these effects~\cite{FP00,BBBCEHLP05,MS06,BBBCS11}.
Therefore, it is not a surprise that two popular approaches to glasses, Mode
Coupling Theory (MCT)~\cite{Go09} and the replica method~\cite{MP09,PZ10}, both agree 
with the Adams--Gibbs
scenario and predict large scale dynamical heterogeneities with a
dynamical correlation length that diverges at the transition to the glass phase.
This qualitative prediction is very interesting, but in order to
make further progresses it would be important to get quantitative
predictions, that can be compared with numerical simulations and with
experiments.

At the mean field level, where both thermodynamic and
dynamic aspects can be solved exactly, it is found that the replica and Mode-Coupling 
approaches are intimately related.
The study of spherical $p$-spin
models, where dynamics is exactly described by a schematic MCT
equation and equilibrium display glassy phenomena related to Replica
Symmetry Breaking (RSB), shows how the glass transition described by MCT
is related to the emergence of metastable states in equilibrium~\cite{CK93,CC05}.
That basic observation, made more then 20 years ago by
Kirkpatrick, Thirumalai and Wolynes~\cite{KTW89},
opened the way to the application of the
mean field theory of spin glasses to the physics 
of supercooled liquids and glasses~\cite{LW07,Ca09,BB11}.
Despite this clear relation at the level of mean field schematic models,
when one tries to apply the mean field theory to realistic models of simple liquids~\cite{MP96,CFP98,MP09,Go09,PZ10}
approximations are mandatory, and because of that the
connection between statics and dynamics becomes more difficult to establish.
It has been shown by Szamel~\cite{Sz10} that under suitable
approximations, similar to the one of MCT, the long time limit of the
MCT equations could be derived from a replicated liquid
theory. Unfortunately this leads to expressions that are not
variational and one cannot get an approximation for the free energy
from the computation. Using instead standard liquid theory approximations
within replica theory~\cite{MP96,CFP98,MP09,PZ10}, one
finds strong discrepancies between predictions from MCT and replicas
which become particularly pronounced in large dimensions~\cite{PZ10,CIPZ11}. 

Besides this consistency problem, 
in finite dimensions one would like to compute the corrections due to fluctuations around the mean field approximation.
When this program is carried out, one finds that there are two important sources 
of corrections to the mean field scenario. The first corrections originate from critical
fluctuations that become important around the glass transition below the upper critical
dimension, as in any standard critical phenomenon~\cite{BB07,FPRR11}.
The second corrections are non-perturbative phenomena related to activated processes.
They can be taken into account by a phenomenological approach,
leading to a number of predictions that are in good agreement with experiment~\cite{LW07};
however, the theoretical foundations of this approach are still controversial~\cite{BB09}
and alternative (but possibly related) phenomenological descriptions of activated relaxation in glasses
have been developed, mostly based on the concept of dynamical facilitation~\cite{KHGGC11}.

In this paper we will only consider critical fluctuations around mean field, so we will not take into account
activated processes. 
Critical fluctuations have been previously described within MCT~\cite{BBMR06,BB07,BBBKMR07a,BBBKMR07b}.
However, field theoretical methods are not yet under complete control in the context of dynamics, and it is therefore
extremely important to set up a static replica field theoretical
description of dynamical heterogeneities, in such a way that well-established 
equilibrium field theory methods such as the renormalization group can be applied to the glass transition problem.
This is what we achieve in this paper. We obtain a low-energy effective action that describes critical fluctuations on approaching
the glass transition, whose coupling constants are obtained directly from the inter-particle
interaction potential using standard liquid theory.
This allows us to compute prefactors to the singular
behavior of physical observables in the mean field approximation, such as
the correlation length or the four-point correlation functions. 
In addition, we show that an important characterization of
dynamics, the MCT exponents, can be obtained within the static replica
framework.
Using the well established HNC approximation of
liquid theory, we perform explicit computations
for Hard and Soft Sphere models and Lennard-Jones
potentials and we obtain good agreement with available numerical data.  
Finally we introduce a quantitative Ginzburg criterion defining a region where perturbative
corrections to mean field theory can be neglected.

\subsection{Dynamical heterogeneities}
In the following we consider a system of $N$ particles in a volume $V$ interacting through a pairwise potential $v(r)$ in a $D$ dimensional space.
The dynamical glass transition is characterized by an (apparent) divergence of the relaxation time of density fluctuations,
that become frozen in the glass phase. 
If $\hat\r(x,t) = \sum_{i=1}^N \d(x-x_i(t))$ is the local density at point $x$ and time $t$ and $\r = \la \hat\r(x,t) \ra$ its equilibrium average, 
the transition can be conveniently characterized using correlation
functions. Consider the density profiles 
at time zero and 
at time $t$, respectively given by 
$\hat\rho(x,0)$ and $\hat\rho(x,t)$. We can define a local similarity measure of these 
configurations as 
\beq\label{ove}
\hat C(r,t) = \int \de x f(x) \hat\r\left(r+\frac{x}2,t\right) \hat\r\left(r-\frac{x}2,0\right)- \r^2 \ ,
\eeq
where $f(x)$ is an arbitrary ``smoothing"
function of the density field with some short range $A$.
In experiments, $f(x)$ could describe the resolution of the detection system
and can be for instance a Gaussian of width $A$. 

Let us call $C(t) =V^{-1} \int \de r \langle \hat C(r,t) \rangle$ the spatially and thermally averaged correlation function.
Typically, on approaching the dynamical glass transition $T_{\rm d}$, 
$C(t)$ displays a two-steps relaxation, with a fast ``$\b$-relaxation'' 
occurring on shorter times down to a ``plateau'',
and a much slower ``$\a$-relaxation'' from the plateau to zero~\cite{Go09}.
Close to the plateau at $C(t) = C_{\rm d}$,
one has $C(t) \sim C_{\rm d} + \AA \, t^{-a}$ in the $\b$-regime.
The departure from the plateau (beginning of $\a$-relaxation)
is described by $C(t) \sim C_{\rm d} - \BB \, t^b$.
One can define the $\a$-relaxation time by $C(\t_\a) = C(0)/e$. 
It displays an apparent power-law divergence at the transition, 
$\t_\a \sim |T-T_{\rm d}|^{-\g}$.
All these behaviors are predicted by MCT~\cite{Go09}.
In low dimensions, a rapid crossover to a different regime dominated by activation 
is observed and the divergence at $T_{\rm d}$ is
avoided; however, the power-law regime is the more robust the higher the dimension~\cite{CIMM10,CIPZ12}
or the longer the range of the interaction~\cite{IM11}.

It is now well established, both theoretically and experimentally, 
that the dynamical slowing is accompanied by growing heterogeneity of the local relaxation, in the sense that the local
correlations $\hat C(r,t)$ display increasingly correlated fluctuations when $T_{\rm d}$ is approached~\cite{FP00,DFGP02,BBBCEHLP05,BBBCS11}.
This can be quantified by introducing the correlation function of $\hat C(r,t)$, i.e. a four-point dynamical correlation
\beq\label{G4}
\begin{split}
&G_4(r,t)=
 \langle 
\hat C(r,t) \hat C(0,t) 
\rangle -\langle 
\hat C(r,t)\rangle\langle \hat C(0,t) 
\rangle 
\end{split}\eeq
The latter decays as $G_4(r,t)\sim \exp(-r / \xi(t))$ with a ``dynamical correlation length'' 
that grows at the end of the $\b$-regime and
has a maximum $\xi = \xi(t\sim \t_\a)$
that also (apparently) diverges as a power-law when $T_{\rm d}$ is approached.

MCT~\cite{Go09} and its extensions~\cite{BBMR06,BB07,BBBKMR07a,BBBKMR07b,Sz08,SF10} 
give precise predictions for the critical exponents.
However, as discussed in the Introduction, this dynamical transition can be also described, at the mean field level, in a static framework.
This has the advantage that calculations are simplified so that the theory can be pushed much forward, in particular by constructing a reduced
field theory and setting up a systematic loop expansion that allows to obtain detailed predictions for the upper critical dimension and the critical
exponents~\cite{FPRR11}. Moreover, very accurate approximations for the static free energy of liquids have been constructed~\cite{Hansen}, and one
can make use of them to obtain quantitative predictions for the physical observables. This is the aim of the rest of this paper.

 \subsection{Connection between replicas and dynamics}
 In the mean field scenario, the dynamical transition of MCT is related to the emergence of a large number of metastable states
 in which the system remains trapped for an infinite time. At long times in the glass phase, the system is able to decorrelate within
 one metastable state. Hence we can write
 \beq
\langle  \hat C(r,t \to\io) \rangle = \int \de x f(x)  \overline{\la \hat\r\left(r+\frac{x}2 \right) \ra_{\rm m} \la \hat\r\left(r-\frac{x}2\right) \ra_{\rm m}}- \r^2 \ ,
 \eeq
 where $\langle \bullet \rangle_{\rm m}$ denotes an average in a metastable state, and the overline denotes an average over the metastable
 states with equilibrium weights.
 
 The dynamical transition can be described in a static framework
by introducing a replicated version of the system~\cite{Mo95,MP96}: 
for every particle we introduce $m-1$ additional particles identical to the first one. 
In this way we obtain $m$ copies of the original system, labeled by $a=1,\ldots,m$.
The interaction potential between two particles belonging to replicas $a,b$ is $v_{ab}(r)$.
We set $v_{aa}(r) = v(r)$,
the original potential, and we fix $v_{ab}(r)$ for $a\neq b$ to be an attractive potential that constrains
the replicas to be in the same metastable state.
Let us now define our basic fields that describe the one and two point density functions
\beq\label{physical_fields}
\begin{split}
&\hat\rho_a(x)=\sum_{i=1}^N\delta(x-x_i^{a}) \ , \\
&\hat\rho_{ab}^{(2)}(x,y)=\hat\rho_a(x)\hat\rho_b(y)-\hat \rho_a(x)\delta_{ab}\delta(x-y) \ .
\end{split}
\eeq

To detect the dynamical transition one has to study the two point correlation functions when
$v_{ab}(r)\to 0$ for $a\neq b$, and in the limit $m\to 1$ 
which reproduces the original model~\cite{Mo95,MP96}.
In this limit, the two-replica correlation function is, for $a\neq b$:
\beq
\langle  \hat C_{ab}(r) \rangle = \int \de x f(x)  \la \hat\r_a\left(r+\frac{x}2 \right) \hat\r_b\left(r-\frac{x}2\right) \ra - \r^2 \ .
\eeq
Because of the limit $v_{ab}(r)\to 0$, the two replicas fall in the same state but are otherwise uncorrelated inside the state,
therefore we obtain $\langle  \hat C_{ab}(r) \rangle = \langle  \hat C(r,t \to\io) \rangle $ which provides the crucial identification
between replicas and dynamics. Similar mappings can be obtained for four-point correlations.

\subsection{Replica field theory for the dynamical transition} 
We introduce for convenience an external field $\nu_a(x)$ (that derives from a space-dependent chemical potential),
in such a way that the density correlation functions
can be obtained by taking the derivative of the free-energy with respect to it~\cite{Hansen}. 
The free energy is defined as the logarithm of the partition function, and
its double Legendre transform defines the Gibbs free energy $\Gamma[\{\rho_a(x)\},\{\rho^{(2)}_{ab}(x,y)\}]$~\cite{Hansen,MH61}:
\beq
\begin{split}
&\G =  \frac 12 \sum_{a,b} \int \de x \de y \left[ \r^{(2)}_{ab}(x,y) \ln \left( \frac{\r^{(2)}_{ab}(x,y)}{\r_a(x) \r_b(y)}
 \right) \right. \\
 & - \left. \r^{(2)}_{ab}(x,y) + \r_a(x) \r_b(y) \right]  + \sum_a \int dx \r_a(x) \left[ \ln \r_a(x) - 1 \right] \\
&+ \hskip-10pt  \sum_{n \ge 3, a_1,\ldots,a_n} \hskip-10pt \frac{(-1)^n}{2n}
\int \de x_1 \cdots \de x_n \, \r_{a_1}(x_1) h_{a_1 a_2}(x_1,x_2) \times \\
& \times \cdots \r_{a_n}(x_n) h_{a_n a_1}(x_n,x_1)  + \G_{\rm 2PI} \ ,
\label{HNC}
\end{split}
\eeq
where $h_{ab}(x,y) = \r^{(2)}_{ab}(x,y)/\r_a(x) \r_b(y) - 1$ and 
$\G_{\rm 2PI}$ is the sum of 2-line irreducible diagrams~\cite{MH61}.
The average values of the fields in Eq.~\eqref{physical_fields}, namely $\overline\rho_a(x)$ and $\overline\rho_{ab}(x,y)$, can be obtained by solving the saddle point equations
\begin{equation}\label{SP}
\left.\frac{\delta \Gamma[\{\rho_a\},\{\rho^{(2)}_{ab}\}]}{\delta \rho_{ab}^{(2)}(x,y)}\right|_{\overline \rho_{ab}(x,y)}=\frac{1}{2}v_{ab}(x,y) \ ,
\end{equation}
and similarly for $\rho_a(x)$.
Here we consider a homogeneous liquid, hence $\rho_a(x)=\rho$. 

We have to assume at this point that a mean field approximation of the free energy is available, that we shall use as
the starting point of our computations.
Within this approximation, 
we want to study the behavior of $\overline\rho_{a\neq b}(x,y)$ in the double limit $m\to 1$ and $v_{a\neq b}\to 0$, which
signals the dynamical transition:
if $T > T_{\rm d}$, then $\overline\rho_{a\neq b}(x,y)=\r^2$
while if $T\leq T_{\rm d}$ a non trivial off-diagonal solution persists in the limit $v_{a\neq b}\to 0$. 
At the mean field level, the appearance 
of the non trivial solution is a bifurcation phenomenon so that, if we come from below 
the transition and we define $\epsilon =T_{\rm d}-T$, we have for $\ee\to 0$:
\begin{equation}\label{offdiagonal}
\overline\rho_{a\neq b}(x,y;\epsilon)= \rho^2\widetilde g(x-y)+2 \rho^2\sqrt\epsilon \ \kappa \  k_0(x-y) \ ,
\end{equation}
where $k_0(x)$ is normalized as $\int \de x \, k_0(x)^2 = 1$, and
$\kappa$ is a constant.
From the saddle point equations~\eqref{SP} we obtain that the Hessian matrix for the off-diagonal elements, \ie for $a\neq b,\ c\neq d$
\begin{equation}
M_{ab;cd}(x_1,x_2;x_3,x_4)=\frac{\delta^2 \Gamma[\{\rho_a\},\{\rho^{(2)}_{ab}\}]}{\delta\rho_{ab}^{(2)}(x_1,x_2)\delta\rho^{(2)}_{cd}(x_3,x_4)}\label{Mass}
\end{equation}
considered as a kernel operator both in standard and replica space
develops a zero mode at $T_{\rm d}$.  This means that if the
transition is approached from below, the fundamental eigenvalue of
this operator is proportional to $\sqrt\epsilon$ due to the
bifurcation-like phenomenology.  Moreover the eigenvector
corresponding to it is $k_0(x-y)$. 

Exploiting the replica symmetry of the saddle point solution Eq.~\eqref{SP}, the most general form of the Hessian matrix is given by
\begin{gather}
M_{ab,cd}(x_1,x_2;x_3,x_4)=M_1\left(\frac{\delta_{ac}\delta_{bd}+\delta_{ad}\delta_{bc}}{2}\right)+\nonumber\\
+M_2\left(\frac{\delta_{ac}+\delta_{ad}+\delta_{bc}+\delta_{bd}}{4}\right)+M_3
\end{gather}
where $M_1$, $M_2$ and $M_3$, depend on $x_1,\ldots, x_4$. 
From this one can show that, because the zero mode $k_0(x-y)$ is independent of the replica indices, 
in the replica limit $m\to 1$ it is an eigenvector of the kernel operator $M_1$.
To study the correlation functions for the fields in Eq.~\eqref{physical_fields} we can produce a 
power series expansion of the Gibbs free energy in terms of the fluctuation of the field $\r^{(2)}_{a\neq b}(x,y)$ 
from its saddle point value. Defining the field $\Delta\rho_{ab}(x,y) =\rho_{ab}^{(2)}(x,y) -\overline\rho_{ab}(x,y)$, 
we can expand the Gibbs free energy up to the third order. It is convenient 
to define $p_i$ and $q_i$ as the momenta conjugated to the half sum 
and the difference of the spatial arguments of 
$\Delta\rho_{ab}(x_i,y_i)$. Using translation invariance 
we write the replica action in Fourier space as
\beq
\begin{split}
& \Gamma[\{\Delta\rho_{ab}\}]=\Gamma[\{\overline\rho_{ab}\}]+ \\
&\frac{1}{2}\sum_{a\neq b,c\neq d}\int \frac{\de p\de q_1\de q_2}{(2\pi)^{3D}}\Delta\rho_{ab}(p,q_1)M^{(p)}_{ab;cd}(q_1,q_2)\Delta\r_{cd}(-p,q_2) +\\
&\frac{1}{6}\sum_{ab;cd;ef}\int \frac{\de p\de p'\de q_1 \de q_2\de q_3}{(2\pi)^{5D}} L_{ab;cd;ef}(p,p';q_1,q_2,q_3) \times \\
&\times\Delta\r_{ab}(p,q_1)\Delta\r_{cd}(p',q_2)\Delta\r_{ef}(-p-p',q_3)\label{Gamma_Delta}
\end{split}
\eeq 
Because of the zero mode of the
Hessian matrix, the connected correlation function of $\Delta\rho_{ab}(x,y)$
shows critical fluctuations at the transition. 

To
make the connection with the dynamical correlation, we define an
overlap function among replicas, $q_{ab}(r)$ as in Eq.~\eqref{ove}
substituting the configurations at time 0 and $t$ by
replicas $a$ and $b$.
We expect that all the critical fluctuations of $q_{ab}(r)$ can be
captured by a projection on the zero mode, leading from
Eq.~\eqref{Gamma_Delta} to an effective action. We can study the
fluctuations of $q_{ab}(r)$ for generic functions $f$, by performing a
Legendre transform of Eq.~\eqref{Gamma_Delta}. However the results are
quite involved and here for clarity we will first consider
the simplest case where $f(x)=k_0(x)$.  Of course, this is not a
practical choice for numerical simulations or experiments because
$k_0$ is quite difficult to measure, however the theoretical
computations are much simpler in this case.  Later on we will show
that any other choice of $f$ leads to the same results for the critical
quantities, and it only affects the prefactor of the correlation
functions.  The projection onto the zero mode can be done
by choosing $\D
\r_{ab}(x,y) = k_0(x-y) \phi_{ab}\left(\frac{x+y}2\right)$ and
substituting this in Eq.~\eqref{Gamma_Delta}. 
The field
$\phi_{ab}(x)$ is the component 
of 
the overlap along the zero mode,
and we perform a perturbative expansion at small momentum $p$.  The
effective replica field theory that arises is equivalent to a
Landau-like gradient expansion along the critical modes:
\beq\label{Gamma_Phi}
\begin{split}
& \Gamma[\{ \phi_{ab}\}]=\frac{1}{2}\int \frac{\de p}{(2\pi)^D}\;\left(
\sum_{a\neq b} (\mu\sqrt{\epsilon}+\sigma p^2) |\phi_{ab}(p)|^2\right.
\\& \left. +m_2 \sum_a\left|\sum_b \phi_{ab}(p) \right|^2 +m_3 \left|\sum_{a\ne b} \phi_{ab}(p) \right|^2 
\right)\\
& +\frac{w_1}{6}\int \sum_{a\neq b\neq c\neq a}\frac{\de p\de p'}{(2\pi)^{2D}} \phi_{ab}(p)\phi_{bc}(p')\phi_{ca}(-p-p') \\
&+\frac{w_2}{6}\int \sum_{a\neq b}\frac{\de p\de p'}{(2\pi)^{2D}}\phi_{ab}(p)\phi_{ab}(p')\phi_{ab}(-p-p') \ ,
\end{split}
\eeq
Eq.~\eqref{Gamma_Phi} is the effective low-energy replica field theory
we will use to compute the critical properties of the system. All its 
coefficients can in principle be computed from the
microscopic details of the systems once an approximation 
for $\Gamma$ is available.  In fact they can be given explicit expressions
as functions of derivatives of 
the Gibbs free energy and of the zero mode, which 
both derive from the interaction potential
(see Supporting Information).

\subsection{Correlation functions, correlation length 
and critical exponents}

The effective replica field theory in Eq.~\eqref{Gamma_Phi} can be
used to compute the MCT parameter $\lambda$.  This quantity is related
to the MCT critical exponents that control the approach to the plateau
by the relation
\begin{equation}
\lambda=\frac{\Gamma(1-a)^2}{\Gamma(1-2a)}=
\frac{\Gamma(1+b)^2}{\Gamma(1+2b)} \ .
\end{equation}
In addition, the exponent that controls the growth of the
relaxation time $\t_\a \sim |T - T_{\rm d }|^{-\g}$ is given by $\g =
1/(2a) + 1/(2b)$.  Although $\l$ is a dynamical parameter, it has
been explicitly shown recently in disordered mean field models and it
can be argued on general ground~\cite{CFLPRR12,PR12},
that this parameter can be related to a ratio of 6 point static correlation
functions computable in the replica field theory that we have just
derived. In this scheme the exponent parameter is given by
\begin{equation}
\lambda=\frac{w_2}{w_1}.
\end{equation}

Moreover, the field theory above can be used at the 
Gaussian level in order to obtain the correlation functions 
of the overlap.
The analysis of the quadratic part of Eq.~\eqref{Gamma_Phi} shows that
the correlation length is controlled by the diagonal part, 
being $m_2$ and $m_3$ finite at the transition. The result is
\begin{equation}\label{xi}
\xi = \xi_0 \epsilon^{-1/4} \ , \hskip1cm
\xi_0=\sqrt{\frac{\sigma}{\mu}} \ .
\end{equation}
and it corresponds to the divergence of the dynamical correlation
length $\xi(t)$ in the $\b$-regime~\cite{BBMR06,BBBKMR07a,BBBKMR07b}.

Moreover we can compute in details the critical behavior of many
possible dynamical four-point functions, that are identified with
different matrix elements of the inverse of the Hessian matrix in
Eq.~\eqref{Mass}, see~\cite{FPRR11}.  Here we give the results for 
the simplest one, the so-called in-state, or thermal susceptibility, 
that is given by \beq\begin{split}
  &G_{\rm th}(r,t)=\mathbf E_0
\left[ \langle 
\hat C(r,t) \hat C(0,t) 
\rangle -\langle 
\hat C(r,t)\rangle\langle \hat C(0,t) 
\rangle 
\right]
\end{split}\eeq
where in the equation above $\mathbf E_0[\cdot]$ has to be intended as the average over the initial positions of the particles,
while $\la \bullet\ra$ is an average over different trajectories
(i.e. over the noise for Langevin dynamics, or over the initial velocities for
Newton dynamics).
In the long time limit, this quantity is one of the critical 
contributions to the $G_4(r,t)$ in Eq.~\eqref{G4}, and it can be 
computed directly from the replica field theory above~\cite{FPRR11}.
Here we had to generalize the calculation of~\cite{FPRR11} to 
take into account the structure of the zero mode and
the presence of the smoothing function $f(x)$. The result is
\begin{equation}\label{Gth}
G_{\rm th}(p)=\frac{G_0 \epsilon^{-1/2}}{1+\xi^2p^2}\ , \ \ \  \ G_0=\frac{1}{\mu}\int \frac{\de q}{(2\pi)^D}f(-q)k_0(q) \ .
\end{equation}
We obtain that the correlation length and its prefactor are not dependent on the function $f(x)$ and always given by Eq.~\eqref{xi}. 
The only dependence on $f(x)$ of the four-point function is in the prefactor $G_0$. 
The full four-point correlation~\eqref{G4} is known to display a doubled
singularity with respect to~\eqref{Gth}. In fact, with the choice
$f(x)=k_0(x)$ one finds
$G_{4}(p) = G_{\rm th}(p) - (m_2+m_3)G_{\rm th}(p)^2$~\cite{FPRR11}.
For generic $f(x)$, the computation of the prefactor is more involved
and will not be presented here.

\subsection{A Ginzburg Criterion}

All the calculations above are based on the assumption that a mean field
approximation of the free energy of the system is given.  From this,
we derive the effective Landau field theory Eq.~\eqref{Gamma_Phi}.
From its coefficients, we extracted all the mean field critical
exponents, as well as microscopic expressions for the prefactors.  Now
we can check whether loop corrections to the effective field theory
affect strongly the mean field predictions, by means of a
Landau-Ginzburg computation.  In other words we want to see whether
the loop corrections to the bare correlation function are small.  
In
principle we should take the field theory derived above, and then we
should compute the first non trivial loop diagrams which give the
first correction to the propagator in replica space.  This computation
is quite involved because we have to deal with replica indices.
However it has been shown in~\cite{FPRR11} that the leading divergent
behavior of the above field theory can be mapped to the one of a
scalar field in a cubic potential with a random field
\beq
\begin{split}
S&(\phi) =\frac{1}{2}\int \de x\,\phi(x)(-\sigma \nabla^2+\mu\sqrt{\epsilon} + \d m(g,\D))\phi(x) \\
&+\frac{g}{6}\int\de x\phi^3(x)+\int\de x ( h_0(x) + \d h(g,\D) )\phi(x).
\end{split}
\eeq
where the random field has zero mean and correlation 
$\overline{h_0(x)h_0(y)}=\Delta \delta(x-y)$,
and the coupling constants are given by
$g=w_2-w_1$ and
$\Delta =-m_2-m_3$.

The terms $\d m(g,\D)$ and $\d h(g,\D)$ are counterterms needed to enforce that the critical point 
is not shifted by loop corrections. By computing the first one-loop diagrams and by imposing 
that the relative correction is small with respect to the bare quantity, we arrive to the following Landau-Ginzburg criterion
\beq
1\gg \textrm{Gi}\, \xi^{8-D}
\eeq
where the (dimensional) Ginzburg number is given by
\beq
\textrm{Gi}=\frac{g^2\Delta}{4(4\pi)^{D/2}}\Gamma\left(4-\frac{D}{2}\right) \ .
\eeq
This computation is correct only below the upper critical dimension $D_{\rm u}=8$. For $D \geq D_{\rm u}$, 
the theory is divergent in the ultraviolet and the Ginzburg number depends on the microscopic details, but the critical
exponents coincide with the mean field ones.
A similar calculation in the framework of MCT has been carried out by Szamel~\cite{Sz12}.

\subsection{Results in the HNC approximation}

Up to now the calculations were very general and the results above hold for any given 
approximation of the replicated
free energy functional that displays the correct mean field glassy phenomenology. 
One of the advantage of our static approach 
is indeed that it can be systematically improved
by considering more accurate approximations of $\G$.

Here we report results obtained from the replicated HNC approach, 
that amounts to neglecting the $\G_{\rm 2PI}$ term in Eq.~\eqref{HNC}, and
has been shown to give the correct glassy phenomenology at the mean field level~\cite{MP96,CFP98}. 
Applying the formulae above, we find that
in the HNC approximation the parameter $\l$ is given by
\begin{equation}\label{lHNC}
\lambda=\frac12 \frac{\frac{1}{\r^4}\int \de x \frac{k_0^3(x)}{\widetilde g^2(x)}}{\frac{1}{\r^3}\int \frac{\de q}{(2\pi)^D}k_0^3(q)\left[1-\r \Delta c(q)\right]^3}
\end{equation}
where $\widetilde g(x)=\overline \r_{a\neq b}(x)/\r^2$, 
$\Delta c(q)=c_{aa}(q)- c_{a\neq b}(q)$,  
and the direct correlation function $c_{ab}(q)$ is related to $h_{ab}(q)$ by the replicated Ornstein-Zernicke relation~\cite{MP96}.
Similar expressions can be obtained for all the other coefficients, see Supporting Information.

To produce concrete numerical results we have solved numerically 
the HNC equations by standard methods~\cite{MP96} 
for a large variety of systems in $D=3$. 
In particular we have considered
\begin{itemize}
\item Hard Spheres (HS): $v(r) = 0$ for $r>r_0$ and $v(r) = \io$ otherwise.
\item Harmonic Spheres (HarmS): $v(r) = \e (r_0-r)^2 \th(r_0-r)$.
\item Soft Spheres (SS-$n$): $v(r) = \e (r_0/r)^{n}$, with $n=6,9,12$.
\item Lennard-Jones (LJ): $v(r) = 4 \e \left[ (r_0/r)^{12} - (r_0/r)^6 \right]$
\item Weeks-Chandler-Andersen (WCA): \\ $v(r) = 4 \e \left[ (r_0/r)^{12} - (r_0/r)^6 + 1/4 \right] \th(r_0 2^{1/6}-r)$ 
\end{itemize}
In all cases we fix units in such a way that $r_0=1$, $\e=1$ and the Boltzmann 
constant $k_{\rm B}=1$.  For
HS and SS, temperature is irrelevant (for SS the only relevant
parameter is a combination of density and temperature, hence we fix
$T=1$ for convenience), and we study the system as a
function of density to determine the glass transition density $\r_{\rm d}$.  
For the other systems, we studied the transition as a function of both density and temperature.

In order to obtain numerically the zero mode we have used the
definition in Eq.~\eqref{offdiagonal}, and estimated it by the
numerical derivative of $\widetilde g(r)$ with respect to $\sqrt{\ee}$
when $\ee\to 0$. A plot of the
zero mode for HS is in Fig.~1. Interestingly we find that
the zero mode
has the same structure in Fourier space as the static structure factor $S(q)$ and
the non-ergodicity parameter $f(q)$, which is the Fourier transform of the long time limit 
of Eq.~\eqref{ove} in the glass phase~\cite{Go09}. 
This finding offers a rationalization
of the common practice of concentrating on momenta of the 
order of the peak of $S(q)$ in the study of glassy relaxation. 

\begin{figure}
\includegraphics[width=.95\columnwidth]{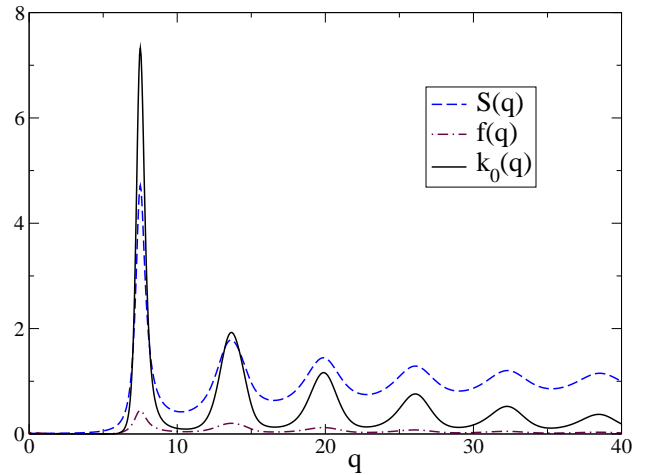}
\caption{
The zero mode $k_0(q)$, the structure factor $S(q)$ and the non-ergodicity factor $f(q)$
for Hard Spheres at the dynamical transition $\r_{\rm d}=1.176$ in the HNC approximation. 
}
\label{fig1}
\end{figure}

From the zero mode we can compute all
the coefficients of the effective action from which we obtain the
physical quantities. 
In particular, we can compute the prefactor $\xi_0$ of the growth of the correlation length and the Ginzburg number.
Moreover, we have computed the prefactor $G_0$ of the
in-state susceptibility Eq.~\eqref{Gth} using a box function 
$f(x) = (2 A)^{-D/2} \prod_{\a=1}^D \theta(A^2-x_\a^2)$ where $\th(x)$ is the Heaviside step function
and $A=0.1 r_0$. All the results are
collected in tables~1 and~2.

The value of $\lambda$ we find is almost the same for all investigated systems but it is not consistent with the result of MCT~\cite{Go09}
and with numerical results for these systems.
Note, moreover, that the location of the critical point predicted by HNC is different from the one of MCT: e.g. for HS, HNC predicts $\r_{\rm d} = 1.169$ while
MCT predicts $\r_{\rm d} = 0.978$~\cite{Go09}. This is an example of the fact, already mentioned in the introduction, that different approximation schemes
lead to different results. 
Another example of this problem is obtained by comparing the results for LJ and WCA at $\r=1.2, 1.4$ (table~2) with MCT and numerical 
data reported in table~1 of Ref.~\cite{BT10}.
The most interesting numerical result is the Ginzburg number. We predict that (perturbative)
corrections to mean field results in $D=3$ should remain small as long as the dynamical correlation length is
smaller than $\sim 1$. Note that a different Ginzburg criterion for the validity of MCT, based on a phenomenological approach, 
has been derived in~\cite{BB09}: the results of that analysis also suggest that corrections to mean field will appear when the correlation
length is $\sim 1$.

Unfortunately, not many data for the critical behavior of
four-point correlations in the $\beta$-regime are available~\cite{SA08,KDS10}. It
would thus be very interesting to get high precision simulation data in the $\beta$-regime.

\subsection{Conclusions}
We have studied in details the replica field theory for the dynamical transition in glasses. 
By using the HNC approximation we have computed many physical observables directly from the microscopic expression of
the interaction potential.
First of all we provided a way to compute the Mode-Coupling exponent parameter $\l$. 
The numerical values obtained are in good agreement with the experimental and numerical estimates. 
Moreover we have computed the prefactor of the correlation length at the transition, 
together with the prefactor of the in-state four-point correlation function. 
Finally we have closed self-consistently our analysis by looking at the loop corrections 
to the mean field quantities in order to produce a Ginzburg criterion that states how close 
we have to be to the dynamical transition in order to see deviations from mean field theory. 
We found that the range currently accessible to numerical simulations in three dimensions is close to the 
point where such corrections should become important.
Of course, non-perturbative corrections (activated processes) are not included in our analysis,
but they are responsible for strong deviations from the MCT regime when the transition is approached.

Our analysis is quite general because it relies only on the assumption that the approximation scheme 
used for the Gibbs free energy shows the correct mean field glassy phenomenology. Hence, it can
in principle be repeated in different approximation schemes in order to go beyond HNC and obtain more accurate
expressions for physical quantities.

\begin{acknowledgments}
We thank F.Caltagirone, P.Charbonneau, U.Ferrari, L.Leuzzi, F.Ricci-Tersenghi, T.Rizzo, R.Stein and G.Szamel for stimulating discussions. H. J. PhD work is
funded by a fondation CFM-JP Aguilar grant. The European Research Council has provided financial support through ERC
grant agreement no.~247328.
\end{acknowledgments}

\section{Supplementary Information}

The Supporting Information text is organized in three parts. In the first one we give a sketch of the line of reasoning that leads to the effective action used to describe the dynamical transition and we give all the expressions for the coefficients of the same action in terms of the interparticle potential. First we do this in a generic framework without specifying the approximation used to compute the Gibbs free energy and then we give the expressions in the HNC case. The second part of the present text is devoted to the Ginzburg criterion: we describe the guidelines of the computation by showing which diagrams have been taken into account to produce the first correction to the bare four-point function that has been given in the main text. The last section contains some details on the numerical calculations and it is useful just to understand how our results can be improved numerically.

\subsection{Coefficients of the replica Gibbs free energy}
As in the main text, we assume that the glassy phenomenology manifests itself in the singular behavior of the off-diagonal field $\r_{a\neq b}(x,y)$ that has a diverging derivative with respect to temperature when the critical point is approached. This implies that the Hessian (or mass) kernel operator develops a zero mode. Actually, we remember here that due to the replica symmetry of the saddle point we have that only one (\emph{i.e.} $M_1$) of the three kernel operators $M_1,\ M_2$, and $M_3$ has a zero mode. This implies that the field $\r^{(2)}$ can be decomposed on the eigenvectors of $M_1$. Because we want to give the expressions for the diverging part of the correlation function, we can simply disregard the excited modes which are finite and take into account only the projection of the dynamical field $\r^{(2)}$ on the zero mode. Practically this is the same as putting to infinity the masses relative to the projections of the dynamical field on the excited states of the kernel operator §$M_1$. By doing this we can produce a gradient expansion for the replicated Gibbs free energy. The simplest way to do this is to impose that the fluctuations of the dynamical field from the saddle point solution are proportional to the zero mode
\begin{equation}
\Delta\r_{ab}(x,y)=\phi_{ab}\left(\frac{x+y}{2}\right)k_0(x-y)\:.
\end{equation}
By doing this, the expressions for the coefficients of the effective action for the critical fluctuations can be computed straightforwardly. Let us consider first the expression for $\sigma$ and $\mu$. They come along in this way. The kernel operator $M_1$ has a ground state eigenvalue $\lambda_0(p)=\mu\sqrt\epsilon +\sigma p^2+O(p^4)$. For small momentum (which means that we look at the correlation of two fluctuations of the dynamical field that are at a very large distance) the expressions for $\mu$ and $\sigma$ can be computed using perturbation theory for the eigenvalue problem for the kernel $M_1$ where the small perturbative parameter is exactly the momentum $p$. The final expressions are given by
\begin{gather}
\mu=\lim_{\epsilon\to 0}\frac{\de }{\de\sqrt\epsilon}\int \frac{\de^D q \de^D k}{(2\pi)^{2D}}k_0(q)M_1^{(p=0)}(q,k)k_0(q)\\
\sigma=\lim_{\epsilon\to 0}\int \frac{\de^D q \de^D k}{(2\pi)^{2D}}k_0(q)\left.\frac{\partial}{\partial p^2}M_1^{(p)}(q,k)\right|_{p=0}k_0(q)
\end{gather}
where the zero mode is supposed to be normalized. In the same spirit the two other masses $m_i$, $i=2,3$, are given by
\begin{equation}
m_i=\lim_{\epsilon\to 0}\int \frac{\de^D q \de^D k}{(2\pi)^{2D}}k_0(q)M_i^{(p=0)}(q,k)k_0(q)
\end{equation}
At this point it is clear how the expressions for the two cubic coefficients $w_1$ and $w_2$ can be obtained; defining
\begin{gather}
L_{ab;cd;ef}(x_1,\ldots,x_6)=\frac{\delta^3\Gamma[\r,\r^{(2)}]}{\delta\r^{(2)}_{ab}(x_1,x_2)\delta\r^{(2)}_{cd}(x_3,x_4)\delta\r^{(2)}_{ef}(x_5,x_6)}
\end{gather}
then they are given by the following expressions
\begin{gather}
w_{1,2}=\int \de^D x_1,\ldots \de^Dx_6 k_0(x_1-x_2)\ldots k_0(x_5-x_6)W_{1,2}
\end{gather}
where
\begin{gather}
W_1=L_{ab,bc,ca}-3L_{ab,ac,bd}+3L_{ac,bc,de}-L_{ab,cd,ef}\\
W_2=\frac{1}{2}L_{ab,ab,ab}-3L_{ab,ab,ac}+\frac{3}{2}L_{ab,ab,cd}+ \nonumber\\
+3L_{ab,ac,bd}+2L_{ab,ac,ad}-6L_{ac,bc,de}+2L_{ab,cd,ef}\:.
\end{gather}
From the expressions for $w_1$ and $w_2$ we can extract the general expression for the exponent parameter $\lambda$. However all the calculation above rely on the assumption that the replicated Gibbs free energy can be computed exactly. This is not possible in the general case and, as we have said in the main text, we have to recast in some given mean field-like approximation which has the correct glassy phenomenology. Here we will give all the expressions above in the HNC approximation where the derivatives of the Gibbs free energy can be computed exactly. The expressions for $\sigma$ and $\mu$ are
\beq
\begin{split}
\mu&=\frac{2\kappa}{\r}\int \frac{\de^D q}{(2\pi)^D}k_0^3(q)\left[1-\r\Delta c(q)\right]-\kappa\int \de^D x\frac{k_0^3(x)}{\r^2\widetilde g^2(x)}\\
\sigma&=\frac{1}{8\rho}\int \frac{\de^D q}{(2\pi)^D}k_0^2(q)\left[\rho\Delta c(q)-1\right]  \times \\ &\times
\left[\left(\Delta c''(q)-\frac{\Delta c'(q)}{q}\right)\cos^2\theta+\frac{\Delta c'(q)}{q}\right] \\ &-\frac{1}{8}\int \frac{\de^D q}{(2\pi)^D}k_0^2(q)\left(\Delta c'(q)\right)^2\cos^2\theta
\end{split}
\eeq
where $\Delta c(q)=c(q)-\tilde c(q)$ is the difference between the diagonal and off-diagonal part of the matrix of the direct correlation functions defined through the Ornstein Zernike equation and $\theta$ is the polar angle in $D$-dimensional polar coordinates. The expressions for the other two mass terms is given by
\beq
\begin{split}
m_2&=-\int \frac{\de^D q}{(2\pi)^D}k_0^2(q)\tilde c(q)\left[\frac{1}{\r}-\Delta c(q)\right]\\
m_3&=\frac 12 \int \frac{\de^D q}{(2\pi)^D}k_0^2(q)\tilde c^2(q)\:.
\end{split}
\eeq
By computing the third derivative of the replicated Gibbs free energy in the HNC approximation we get the expression for $w_1$ and $w_2$:
\beq
\begin{split}
w_1&=-\frac{1}{8 \r^3}\int \frac{\de^D q}{(2\pi)^D}k_0^3(q)\tilde c(q)\left[1-\r \Delta c(q)\right]^3\\
w_2&=- \frac{1}{16 \r^4} \int \de^D x\frac{k_0^3(x)}{\tilde g^2(x)}\:.
\end{split}\:.
\eeq

\subsection{Ginzburg Criterion}
In this section we give a guideline for the computation of the Ginzburg Criterion. In the main text we have said that at the dynamical point where the number of replicas goes to one, the leading behavior of the correlation functions of the two points function $\r^{(2)}$ can be computed using a field theory for a scalar quantity described by a cubic potential in a random field. This observation simplify a lot the loop expansion because it does not involve replica indices that complicate the perturbative analysis. With reference to the action defined in Eq.~[18] of the main text, we can give a perturbative expression for the two point function of the field $\phi(x)$. The bare propagator is given as usual by $G_0^{-1}(p)=\sigma p^2+\mu\sqrt\epsilon +\delta m$. To obtain the two point function it is quite useful to write down the generating functional of the correlation functions $W[J]=\ln Z[J]$ where we can put $J(x)=h_0(x)+\delta h$ and $h_0$ is an external field that can be used to extract the correlation function by taking the derivative with respect to it. Introducing the following diagrammatic notation
\beq
J(x)=\begin{picture}(10,15)(-5,-2)
\SetColor{Black}
\SetWidth{1}
\CCirc(0,0){2}{Black}{White}
\end{picture}
\ \ \ \ \ \ \ 
h_0(x)=\begin{picture}(10,15)(-5,-2)
\SetColor{Black}
\SetWidth{1}
\CCirc(0,0){2}{Black}{Black}
\end{picture}
\ \ \ \ \ \ \ 
\d h(g,\Delta)=\begin{picture}(10,15)(-5,-2)
\SetColor{Black}
\SetWidth{1}
\CCirc(0,0){2}{Red}{Red}
\end{picture}
\eeq
we have that
\beq
\overline{\la \phi(x) \ra} = 
\begin{picture}(30,60)(-15,-2)
\SetColor{Black}
\SetWidth{1}
\Line(0,0)(0,20)
\CCirc(0,20){2}{Red}{Red}
\end{picture}
+ 
\begin{picture}(30,60)(-15,-2)
\SetColor{Black}
\SetWidth{1}
\Line(0,0)(0,20)
\CArc(0,30)(10,0,360)
\CCirc(0,40){2}{Blue}{Blue}
\end{picture}
+
\begin{picture}(30,60)(-15,-2)
\SetColor{Black}
\SetWidth{1}
\Line(0,0)(0,20)
\CArc(0,30)(10,0,360)
\end{picture}\:.
\eeq
We impose that the critical point is not shifted by the perturbative terms so we want also that $\overline{\langle \phi(x)\rangle}=0$ from which we see that the counterterm $\delta h$ is of order $g$. Now let us look at the one loop correction to the propagator. Using the fact that the expectation value of $\phi$ is zero we obtain
\begin{align*}
\overline{\la \phi(x) \phi(y) \ra} &= G_0(x-y) + 
\begin{picture}(50,40)(-25,-2)
\SetColor{Black}
\SetWidth{1}
\Line(-25,0)(25,0)
\CArc(0,0)(15,0,180)
\CCirc(0,15){2}{Blue}{Blue}
\end{picture} +
\begin{picture}(50,40)(-25,-2)
\SetColor{Black}
\SetWidth{1}
\Line(-25,0)(25,0)
\CArc(0,0)(15,0,180)
\end{picture}+\ldots
\end{align*}
We are interested in the most infrared divergent diagrams (in the limit where $T\to T_d$). This means that we can neglect the second diagram, and we can consider only the first one (this is exactly what happens in the perturbative expansion of the Random Field Ising Model). The inverse of the renormalized susceptibility reads
\begin{gather}
m^2_R = G^{-1}(p=0) = m_0^2 + \d m -  \frac{\D g^2}{2(2\pi)^D}\int^\L\de^D q \frac1{(\sigma q^2 + m_0^2)^3}
\end{gather}
where $m_0^2=\m \sqrt{\epsilon}$. By taking the derivative with respect to $m_0^2$ we obtain
\beq
\frac{d m^2_R}{d m^2_0} = 1  +3   \frac{\D g^2}{2(2\pi)^D}\int^\L\de^D q \frac1{(\sigma q^2 + m_0^2)^4}\:.
\eeq
By imposing that the second term on the right hand side is smaller than 1 and by computing the loop integral we get the expressions [19] and [20] of the main text.

\subsection{Details on the numerics}
To produce the numerical values collected in the tables, we have solved numerically the HNC equations in three dimensions. This is a quite easy task 
because such equations can be solved by an iterative Picard scheme. 
However the solution requires the use of Fourier transforms. 
Working in spherical coordinates thanks to the rotational invariance of the system,
we have two natural cutoffs. The first one fixes the maximal distance $L$ (infrared cutoff), hence we only keep $g(r)$ for $0 \leq r \leq L$.
The other one is related to the precision with which we measure the position of the particles (utraviolet cutoff): the possible values of $r$ are discretized 
in such a way that in the unit interval there are $N$ equi-spaced possible positions so that the precision is $1/N$.
The data presented in the tables is relative to the larger cutoffs that we have. In particular, the infrared cutoff is fixed to $L=16$ where the unit distance is the diameter of the particles or the interaction range of the potential. The ultraviolet cutoff is fixed at $N=256$. A remark has to be done on the way we computed the critical point and the zero mode. In fact to observe the correct $\sqrt\epsilon$ behavior of the off-diagonal solution, we need to be quite close to the critical point because otherwise this behavior is hidden by the subleading $\epsilon$ behavior. To give a precise estimate of the critical point we have collected a sequence of solutions of the HNC equation varying the temperature or the density, depending on the case under study, and we have fitted these data with a $\sqrt\epsilon$ behavior. Once we have identified the critical point we have computed the the zero mode using directly the definition given by Eq.~[8] of the main text.

\end{article}

\begin{table*}
\caption{
Numerical values of the coefficients of the effective action and the physical quantities from the HNC approximation. 
For each potential, lengths are given in units of $r_0$ and energies in units
of $\e$, with $k_B=1$. 
Data at fixed temperature, using density as a control parameter with $\ee = \r_{\rm d} - \r$.
}
\label{tab1}
\begin{tabular}{|cc|ccccccc|cccc|}
\hline
System & $T$ & $\rho_{\rm d}$ & $-w_1$ & $-w_2$ & $m_2$ & $m_3$ & $\sigma$ & $\mu$ & $\lambda$ & $\xi_0$ & $G_0$ & Gi   \\
\hline
SS-6  & 1 & 6.691 & 3.88$\cdot10^{-6}$ & 1.35$\cdot10^{-6}$  & -0.000925 & 0.000110 & 0.000195 & 0.000525 & 0.348 & 0.601  & 224  & 0.0267 \\
SS-9  & 1 & 2.912 & 0.0000772  & 0.0000272    & -0.00539  & 0.000633  & 0.00163  & 0.00543  & 0.353 & 0.548  & 34.3 & 0.0125\\
SS-12 & 1 & 2.057 & 0.000275  & 0.0000973    & -0.0116  & 0.00132  & 0.00378  & 0.0152  & 0.354 & 0.498  & 14.2 & 0.0118\\
LJ  & 0.7  & 1.407 & 0.00106  & 0.000376    & -0.0258  & 0.00290  & 0.00989   & 0.0414  & 0.355 &0.489   & 6.00 & 0.00833 \\
HarmS & $10^{-3}$ & 1.336  &   0.00129 & 0.000465 &    -0.0336 & 0.00343 & 0.00772 & 0.0779  & 0.359  & 0.315 & 2.82 & 0.0434\\
HarmS & $10^{-4}$ &  1.196   &  0.00165 & 0.000622 &  -0.0403 & 0.00386 & 0.00819 &  0.109 & 0.378   & 0.274 & 1.69 & 0.0632\\
HarmS & $10^{-5}$  & 1.170  &   0.00174 & 0.000663 &  -0.0416 & 0.00395 & 0.00845  & 0.109 & 0.382   & 0.278 & 1.66 & 0.0635\\
HS &  0   &  1.169  &   0.00174 & 0.000664 &  -0.0418 & 0.00397 & 0.00847 &  0.108 & 0.381 & 0.280 & 1.67 & 0.0639\\
\hline
\end{tabular}
\end{table*}

\begin{table*}
\caption{
Same as table 1, but here the data
are at fixed density, using temperature as a control parameter with $\ee = T_{\rm d} - T$.
}
\label{tab2}
\begin{tabular}{|cc|ccccccc|cccc|}
\hline
System & $\r$ & $T_{\rm d}$ & $-w_1$ & $-w_2$ & $m_2$ & $m_3$ & $\sigma$ & $\mu$ & $\lambda$ & $\xi_0$ & $G_0$ & Gi   \\
\hline
LJ  & 1.2 & 0.336 & 0.00186 & 0.000663 & -0.0361 & 0.00403 & 0.0147 & 0.0572 & 0.356 & 0.507 & 4.56 & 0.00730\\
LJ & 1.27 & 0.438 & 0.00153 & 0.000541 & -0.0321 & 0.00370 & 0.0128 & 0.0447 & 0.353 & 0.536 & 5.74 & 0.00771\\
LJ  & 1.4 & 0.684 & 0.00108 & 0.000383 & -0.0260 & 0.00293 & 0.0100 & 0.0292 & 0.355 & 0.586 & 8.52 & 0.00825\\
WCA & 1.2 & 0.325 & 0.00195 & 0.000686 & -0.0389 & 0.00426 & 0.0133 & 0.0607 & 0.351 & 0.467 & 4.37 & 0.0134\\
WCA  & 1.4  & 0.692 & 0.00111 & 0.000388 & -0.0270 & 0.00301 & 0.00966 & 0.0291 & 0.350 & 0.576 & 8.67 & 0.0106\\
\hline
\end{tabular}
\end{table*}

\end{document}